\documentclass[aps,rsi,floatfix,twocolumn]{revtex4-1}
\usepackage{times}
\usepackage{graphicx}
\usepackage{amsmath}
\usepackage{amssymb}
\usepackage{url}

\begin{document}

\title{Ultrafast laser micro-nano structuring of transparent materials with high aspect ratio}
\author{Francois Courvoisier\\ 
FEMTO-ST Institute, Univ. Bourgogne Franche-Comt\'e, CNRS,\\
15B avenue des Montboucons, 25030 Besan\c con, cedex, France\\ \email{francois.courvoisier@femto-st.fr}
\vspace{1cm}\\
This is a pre-peer-review, pre-copyedit version of a book chapter published by Springer Nature. The final authenticated version is available online at: \\ 
\url{https://doi.org/10.1007/978-3-319-69537-2_33-1}\\
 \vspace{1cm}
}
%
\begin{abstract}
Ultrafast lasers are ideal tools to process transparent materials because they spatially confine the deposition of laser energy within the material's bulk via nonlinear photoionization processes. Nonlinear propagation and filamentation were initially regarded as deleterious effects. But in the last decade, they turned out to be benefits to control energy deposition over long distances. These effects create very high aspect ratio structures which have found a number of important applications, particularly for glass separation with non-ablative techniques. This chapter reviews the developments of in-volume ultrafast laser processing of transparent materials. We discuss the basic physics of the processes, characterization means, filamentation of Gaussian and Bessel beams and provide an overview of present applications.
\end{abstract}
\maketitle


\section*{Introduction}
\label{sec:Introduction}

Ultrafast lasers were very early recognized as a powerful tool to process transparent materials. This is because the transfer of laser energy to the solid starts from nonlinear absorption by ionization. Nonlinear absorption is very efficient only around the focal point of the beam. This way, absorption can be completely negligible on the surface of a dielectric while a large amount of the input pulse energy is absorbed within the bulk of the material. After a number of different physical processes \cite{Mao2004, Gattass2008}, this eventually yields a modification of the bulk material that can range from a simple index change to the formation of an empty cavity. Longer laser pulses, such as nanosecond pulses, can also be used to process transparent materials, but in this case, absorption of the laser pulse initiates from defects. This  is random and poorly reproducible from shot to shot. In contrast, ultrafast pulses, {\it i.e.} sub $\sim$10~ps pulses, modify materials with high reproducibility and with very reduced heat affected zone.

In the last decade, structuring with high aspect ratio has emerged as a new field in ultrafast laser processing of transparent materials. In early works, most of the attention was focused on creating extremely localized damages in 3 dimensions (index change, nanogratings, nano-voids) with high numerical aperture illumination. Nonlinear propagation effects, in the general sense "filamentation", were regarded as highly detrimental. However, it progressively appeared that filamentation of Gaussian and shaped beams, could create  extended modifications along the propagation direction. This was even possible in single shot illumination regime. Controlling these phenomena is a great challenge but allows for creating modifications that are impossible to generate with conventional point-by-point processing. These propagation regimes create high aspect ratio modifications: their length that is much longer than their typical diameter.

High aspect ratio processing is required for a number of applications such as microfluidics, photonics or material separation. As we will describe in section \ref{sec:applications},  an important field of application is glass separation with a non-ablative technique. This technique is based on generating a high aspect ratio structure in the bulk of a transparent brittle material, organized on a plane, which allows the material to cleave along this weakened plane. The process is high speed  and is obviously very well suited to mass fabrication of cover glass for touchscreens and consumer electronics.

However, structuring with high aspect ratio is challenging because the propagation of an intense ultrafast pulses inside a transparent solid is nonlinear by essence. This chapter is intended as a guide in this field of research and technology. We will first briefly review the phenomena occurring during high-intensity pulse propagation in dielectrics as well as the means to characterize plasma formation inside their bulk. We will point out the specificity of bulk propagation and associated numerical simulations. In section \ref{sec:multishot}, a review of high aspect ratio processing in the multiple shot regime will point out the difficulties to face with bulk processing. In contrast, single shot void formation seems more simple and faster. In the related section \ref{sec:void}, we will review the main results which open new routes for laser processing of transparent materials. Increasing the length of the empty cavities can be done with filamentation of Gaussian beams, which we will review in section \ref{sec:filamentation}. But due to the nonlinearities, this process is difficult to predict or scale. In section \ref{sec:Bessel}, we will show that a specific class of beam shapes allows for seeding filamentation such that the propagation is free from distortion in space and time and is therefore much more predictable. This beam shape is the zeroth-order Bessel beam. It induces high aspect ratio index modifications, nanochannels or nano-voids in the bulk of a number of transparent materials. In section \ref{sec:applications}, we will review the applications of both filamentation of Gaussian-like beams and of Bessel beams.  These are numerous and we will describe the technologies as a function of the processed material (glass, sapphire, diamond, or silicon).

\newpage
\section{Ultrafast phenomena and nonlinear propagation in transparent materials}
\label{sec:phenomena}

In this section, we will briefly review the main physical phenomena occurring during high intensity ultrashort laser pulse propagation in the bulk of transparent materials. Our objective is to point out that there are a number of differences in the experimental techniques and numerical modelling in comparison with surface ablation, which has been described in detail in the preceding chapters. 

In brief, the physical sequence of material modification by an ultrafast laser pulse can be split in two main steps: first, nonlinear propagation generates a plasma of free electrons and holes via nonlinear ionization. This is the energy deposition step, which is terminated at the end of the laser pulse, typically in the sub-picosecond range. Then, the second step is the relaxation. It involves a sequence of different physical phenomena extending up to microsecond scale. These phenomena are identical to the ones occurring at the surface of dielectrics. We will therefore focus on the first step, where the nonlinear propagation plays a determinant role for the modifications of the material.

For the numerical modelling, the propagation in the bulk of transparent materials imposes a number of additional constraints in comparison with surface ablation. First, the propagation distances considered in bulk processing of materials are orders of magnitude longer than those simulated for surface ablation. Second, while surface ablation modeling can sometimes be reduced to 0 or 1 dimension, bulk processing requires at least models in two dimensions. We will emphasize in the following how the physics of optical breakdown over long distances can be simulated with reasonably powerful computers. As for experimental characterizations, specific techniques have been developed to characterize plasma formation within the bulk of the materials. We will describe them in the second part of this section.

\subsection{Linear and Kerr effects}

Propagation in transparent materials is determined by several linear and nonlinear effects. As for the linear contributions, the pulse shape is affected by diffraction, dispersion and aberrations such as chromatism or spherical aberration\index{spherical aberration}. Diffraction and aberrations are important effects which explain a number of experimental results \cite{Song2008}.
We note that dispersion in the material can be generally safely neglected because  thicknesses of  materials on the order of a few millimeters only weakly affect ultrashort pulses of duration longer than 100~fs. 

Kerr\index{Kerr effect} effect mainly contributes to the transient index of refraction, as well as for the self steepening of the pulse. Cross phase modulation effects have usually a negligible impact on the pulse intensity. For sufficiently long pulse duration, Raman contribution to Kerr effect can be included. Kerr self-focusing \index{self-focusing} shifts backwards the focusing point as the peak intensity increases.

\subsection{Nonlinear absorption, plasma absorption and plasma defocusing }

The interaction between a laser pulse and a photo-excited solid dielectric is threefold. i) Nonlinear absorption\index{nonlinear absorption} occurs because of high intensity field. The excited electrons interact ii) with the field and iii) in the field via collisions\index{collisions}. 

 From first principles, description of nonlinear absorption and interaction with the laser pulse should be based on a quantum model of the band system in the periodic high electric field \cite{Otobe2008}.
Despite the number of advances in theoretical chemistry, it is still a challenge to accurately describe the ground state of solids such as sapphire, and it is obviously even more difficult for amorphous solids like fused silica. Transition rates between excited states are mostly out of experimental reach, yet a number of highly excited states should be taken into account \cite{Barilleau2016}. Time-Domain Density Functional Theory (TD-DFT) can model the high-field ionization, but collisional absorption is still difficult to describe in this framework \cite{Yabana2012}. In the framework of numerical simulation of laser pulse propagation of several 100's fs over a few millimeters of propagation, these approaches are computationally too demanding.

In first approximation, the excited electrons in the upper bands can be considered as free electrons with effective masses\index{effective mass} (parabolic band). This is why in the rest of the chapter, the excited electrons will be referred to as "free electrons", and nonlinear transitions to the excited states will be referred to as nonlinear ionization\index{ionization}. Thus, the ionization phenomena in dielectrics are described in similar way as ionization of atoms. The modeling of pulse propagation will therefore follow the work that was developed for filamentation in air.

The description of Keldysh\index{Keldysh model} for ionization is computationally efficient. In this framework, multiphoton and tunnel ionization are asymptotic behaviors \cite{Sudrie2002, Couairon2005}. In basic models, the distribution of the free electrons\index{free electrons} in the excited levels is neglected and the free electrons are described only via the number density $\rho$\index{plasma density}. More refined models insert additional spectroscopic levels to describe the energy distribution of the free electrons. This is the multiple rate equations model\index{multiple rate equations} (MRE) \cite{Rethfeld2006}.

The interaction of the laser pulse with the plasma is twofold: i) absorption by the free-electron gas excites the free-electrons to upper levels. When the energy of the free electrons is sufficiently large, impact ionization occurs and the free-electron density increases. ii) the presence of plasma effectively reduces the effective index of refraction, yielding defocusing effect on the laser pulse.

 Drude model\index{Drude model} can be used to efficiently describe this interaction. The plasma conductivity $\sigma (\omega)$ is derived from the plasma number density $\rho$. The plasma can be described as a contribution to the complex permittivity. A frequency-dependent description is valid as long as the number density does not vary too fast in time. Meanwhile, the evolution of the free-electron distribution and impact ionization effects can be either described in the MRE model or by simply considering that every K photons absorbed by the plasma will contribute to the ionization of an additional free electron (see Equation \ref{eq:plasmaeq}).

In detail, the \index{plasma susceptibility} is:

\begin{equation}
    \chi(\omega) = -\frac{\rho e^2}{\varepsilon_0 m}
\end{equation}

\noindent with $\varepsilon _0$ the vacuum permittivity, $e$ the unsigned electron charge, $m$ the effective electron mass.

If the plasma-free permittivity of the medium is $\varepsilon_{SiO_2}$, the combined permittivity\index{plasma permittivity} of the medium and plasma reads \cite{Mao2004}:  

\begin{equation}
\label{eq:epsilon_of_omega}
    \varepsilon(\omega) = \varepsilon_{SiO_2}(\omega)- \omega_p^2 \bigg[ \frac{\tau_c^2}{1+\omega^2\tau_c^2}-i \frac{\tau_c^2}{\omega\tau (1+\omega^2\tau_c^2)}\bigg] =n^2(\omega)
\end{equation}

\noindent $\omega_p=\sqrt{\frac{e^2 \rho}{\varepsilon_0 m}}$ is the plasma frequency and $n$ the complex index of refraction of the medium with the plasma. We see in Equation \ref{eq:epsilon_of_omega} that plasma contribution reduces the permittivity, and therefore reduces the index of refraction. This is why plasma defocuses an incoming laser pulse.

The collision time\index{collision time} $\tau_c$ is a parameter that should be derived from the free-electron plasma density and distribution in momentum space (or temperature for a Maxwellian distribution). In practice, the collision time is usually considered as fixed with  values typically ranging from 0.5~fs (dense plasmas \cite{Velpula2016}) to $\sim$10~fs (low density plasmas \cite{Sudrie2002}).
 Yet imperfect, this model still describes with reasonable accuracy the absorption of the plasma and the change of the local permittivity due to the presence of plasma \cite{Gamaly2011}.
 
Finally, other mechanisms such as recombination, Auger decay do exist and can also be included in models depending on computational capability.

\subsection{Numerical simulations of pulse propagation in transparent dielectrics}

A number of different physical models for ultrafast laser pulse propagation have been developed. Here, we discuss a basic model of propagation to provide the reader a first view on how the different mechanisms described above can be numerically simulated. More detailed models can be found for instance in references \cite{Couairon2011,Bulgakova2014}.
An early model is based on solving simultaneously a NonLinear Schr\"odinger Equation (NLSE\index{NLSE}) and a rate equation for the plasma density \cite{Feit1974,Tzortzakis2001,Sudrie2002,Wu2002,Couairon2005,Winkler2006}. The following NLSE is derived from Maxwell's equations using a scalar, paraxial, unidirectional model for the field envelope $A(x,y,z,\tau)$ describing the laser pulse  with central frequency $\omega_0$ with a temporal reference frame defined by $\tau = t-z/v_g$. $t$ and $z$ are time and propagation distance in the laboratory reference frame and $v_g =c/n_{SiO_2}$ the group velocity.

The NLSE reads \index{NLSE} :

\begin{multline}
    \label{eq:NLSE}
    \frac{\partial A}{\partial z} = \frac{i}{2k}\bigg( \frac{\partial^2}{\partial r^2}+\frac{1}{r}\frac{\partial}{\partial r}\bigg) A-\frac{ik''}{2}\frac{\partial^2 A}{\partial^2\tau^2}+ik_0 n_2 |A|^2A \\
    -\frac{\sigma}{2}(1+i\omega_0\tau_c)\rho A-\frac{W_{PI}(|A|) U_{gap}}{2|A|^2}A
\end{multline}

\noindent and it has to be solved together with the plasma equation:
\begin{equation}
      \label{eq:plasmaeq}
\frac{\partial \rho}{\partial t} = \bigg(W_{PI}(|A|)+ \frac{\sigma}{U_{gap}}\bigg) \bigg(1-\frac{\rho}{\rho_{nt}}\bigg) -\frac{\rho}{\tau_t}
\end{equation}
\noindent where $k=n_0k_0$ is the wavevector inside the medium of refractive index $n_0=\sqrt{\varepsilon_{SiO_2}}$, and $k_0$ the wavevector in vacuum, $k'' = \partial^2 k(\omega) /\partial \omega ^2 |_{\omega=\omega_0}$ is the group velocity dispersion coefficient, $U_{gap}$ is the bandgap of the dielectric medium, $n_2$ is the nonlinear Kerr index, $W_{PI}$ is the nonlinear photoionization rate. $\sigma$ is the plasma conductivity evaluated by Drude model:

\begin{equation}
    \sigma(\omega) = \frac{k\omega_0 e^2 \tau_c}{n_0^2 \varepsilon_0 m \omega_0^2 (1+\omega_0^2\tau_c^2)}
\end{equation}

The first term in equation \ref{eq:NLSE} corresponds to diffraction, the second to dispersion, the third is Kerr effect, the fourth is plasma absorption (real part) and plasma defocusing (imaginary part) and the last one is nonlinear absorption.

 NLSE is usually numerically solved via a split-step algorithm to calculate the pulse shape in $(x,y,t)$ at each point. Simultaneously, the plasma rate equation is solved for the free electron number density $\rho(x,y,z,t)$. Multiple rate equations can be added to describe more accurately the avalanche process \cite{Rethfeld2006}. In addition, for pulses on the order of a several 100's fs to some ps, the transient generation of self trapped excitons (STEs), at timescales of $\sim$150~fs \cite{Guizard1996,Mao2004}, as well as structural defects left by previous pulses, can be described by including the rate equations for new spectroscopic levels  \cite{Bulgakova2014}. 
 
 We note that the terms $(1-\rho / \rho_{nt})$ describe the saturation because of the finite number of available electrons to promote to the conduction band. In practice the number density $\rho_nt$ is estimated by the density of states (typically $2.2\times 10^{22}$ cm $^{-3}$ for fused silica). 
\begin{figure}
    \centering
    \includegraphics[width =\columnwidth]{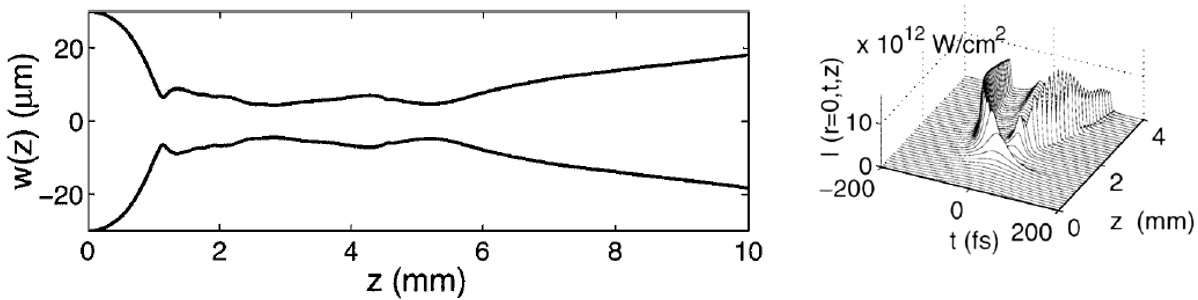}
    \caption{Simulation of ultrafast IR pulse propagating in fused silica. (left) Evolution of the filament diameter along the propagation (right) evolution of the pulse temporal profile on the first 4~mm of propagation. Reprinted figure with permission from \cite{Tzortzakis2001} with courtesy of A. Couairon. Copyright (2001) by the American Physical Society. }
    \label{fig:Simul_filament}
\end{figure}
 Figure \ref{fig:Simul_filament} shows such simulation result for the evolution of the beam diameter and the pulse temporal profile. The temporal distorsion of the pulse is particulary apparent.

Finally, we note that there is no straightforward link between the plasma density and the final modification in the transparent material. This is because a number of physical effects occur after the energy deposition: electron-electron scattering, electron-phonon scattering, recombinations, structural changes, phase changes, shockwaves, thermal diffusion, etc... \cite{Gattass2008}. Void formation occurs approximately when the plasma density approaches the critical plasma density, but this is a crude estimate \cite{Papazoglou2011, Gamaly2006}. Several other parameters have been used to predict the threshold for melting or vaporisation \cite{Grehn2014,Bulgakova2015}. 


\subsection{Experimental diagnostics}
 Experimental characterization is crucial to understand the physics and to identify the regimes in which the modifications are created. Here, we review experimental diagnostics so as to make the reader aware of the potential limitations of the techniques and of the conclusions drawn out of the results obtained.

\subsubsection{Post-mortem diagnostics} "Post-mortem" diagnostics refer to characterizations performed well after the photo-induced phenomena have relaxed. In-bulk material modifications can be characterized only by optical microscopy, including phase contrast, polarized microscopy, Raman techniques. Optical characterization has however a poor spatial resolution ($\sim$0.5~$\mu$m in best cases, depending on probe wavelength and numerical aperture \index{numerical aperture} of the imaging). In addition, spherical aberration\index{spherical aberration} must be compensated to reach the high spatial resolution. In most cases, the sub-micron structures described in this chapter are not resolved by optical means. For higher resolution, only destructive characterization means are available. Mechanical cleavage, polishing or Focused Ion Beam (FIB) milling are used to provide physical access for Scanning Electron Microscopy (SEM). These techniques are extremely delicate because the processing technique should not affect the nano structure itself (filling of empty void by particles, modification of a channel by FIB, "curtain effect", etc). 

\subsubsection{Plasma dynamics characterization}

Plasma dynamics is characterized by pump-probe measurement of the transient distribution of refractive index change. A number of different techniques have been implemented to measure this index change. Several of them are adaptations of techniques initially developed to characterize plasma plumes expanding in vacuum from the surface of solids.

\begin{itemize}
    \item {\it Shadowgraphy} \index{shadowgraphy}is based on transversely illuminating the plasma with a probe pulse. In this case, the complex refractive index change $\Delta n$ is estimated from the transmission $T$ of the probe through the plasma of thickness $L$: $T(x,z)=exp \left[ -4\pi \texttt{Im}(\Delta n) L/\lambda_p \right]$ where $x$ and $z$ are the spatial coordinates in transverse and longitudinal directions respectively, and $\lambda_p$ the central wavelength of the probe \cite{Papazoglou2007,Grossmann2016}. This still requires to estimate a priori the plasma thickness $L$, and to assume that the probe propagation through the plasma is perfectly straight ({\it ie} diffraction effects negligible). Recently, a new tomography \index{tomography}approach was developed to enable the retrieval of the 3D distribution of the extinction coefficient. This approach removes the need to assume the value of the thickness $L$. This is based on multiple shadowgraphy experiments where the beam is rotated around the optical axis between each illumination \cite{Bergner2018}.
    
   Spectrally resolved shadowgraphy is a powerful technique providing access to laser deposited energy under certain approximations \cite{Minardi2014}. For instance, Hashimoto {\it et al} used time resolved  micro-Raman spectroscopy to determine the evolution of temperature distribution after ultrafast excitation of glass \cite{Hashimoto2015}.
 
    \item {\it Pump-probe interferometry} \index{pump-probe measurement} is a technique that retrieves amplitude and phase variations. Depending on implementation, simplified versions of the setup provide access only to the phase measurement, hence to the real part of the refractive index change. The interferometry can be performed with a reference signal that does not cross the interaction medium. Spectral interferometry \index{spectral interferometry}is a technique where 2 probe pulses interfere within a spectrometer, as shown in Figure \ref{fig:SpectralInterferometry}. The reference pulse passes through the medium before the pump, and the second probe records amplitude and phase change with a variable delay with the pump, but fixed delay with the reference. Amplitude and phase can be retrieved from the spectral fringes. This technique is extremely precise, yet it is restricted to characterize a single spatial dimension \cite{Mao2004}. 

    \begin{figure}
    \includegraphics[width=0.5\columnwidth]{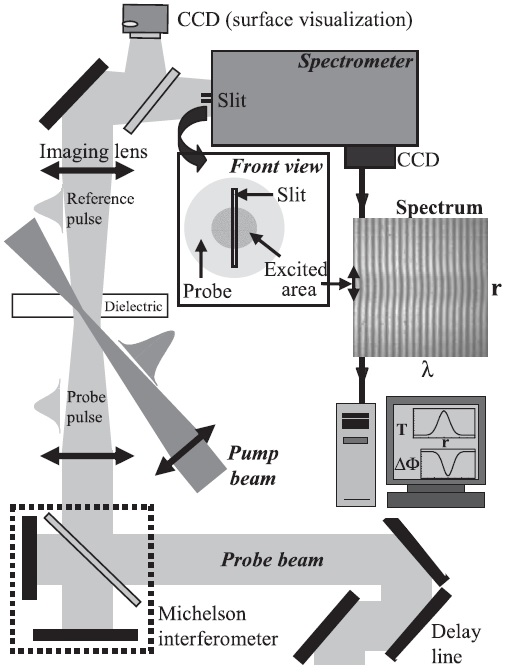}
    \caption{ Example of pump-probe spectral interferometry setup.Reprinted figure with permission from \cite{Mao2004} with courtesy of Prof. S. S. Mao. Copyright (2004) by Springer Nature.}
    \label{fig:SpectralInterferometry}
\end{figure}

    It is also possible to use the interference between the probe wave and the scattered wave emitted by the plasma to characterize the plasma density with holography \cite{Papazoglou2008,Papazoglou2014}. 
    This provides quantitative measurements of phase and amplitude.
    
    Again, despite quantitative measurements can be performed, one must keep in mind that the characterization is convoluted by the optical response function of the imaging apparatus. This actually imposes a severe constraint on the effective spatial resolution of the measurements. After retrieving the distribution of complex index of refraction, Drude model is used to link index change with plasma density, following Eq. \eqref{eq:epsilon_of_omega}.

    \item {\it Two-color probing}. The retrieval of the plasma density from the index change distribution, using Eq. \eqref{eq:epsilon_of_omega} requires the assumption on the collision time $\tau_c$. Repeating the probe measurement with another probe wavelength removes the ambiguity on $\tau_c$ \cite{Velpula2016}. 

 \item {\it Phase contrast microscopy} records images that are proportional to the variation of index of refraction \cite{Zernike1942}. This does not require a second reference probe, and makes straightforward to image the sign of index variations associated to densification or material expansion \cite{Mermillod-Blondin2011}.
\end{itemize}

\subsubsection{Characterization of the pulse distribution }

The pump pulse can be also characterized after its interaction with the solid. This can be easily quantitatively compared to numerical simulation results. This has been performed as spatially resolved crosscorrelation \cite{Grazuleviciute2015}. 
To provide the evolution of the spatiotemporal dynamics along the propagation, the pulse has to be measured at intermediate propagation distances. This is however possible only if the nonlinear propagation stops. This is feasible for instance if the medium has a variable length, because further propagation in air is linear for sufficiently low pulse energies. Jarnac {\it et al} have used a variable size water cuvette \cite{Jarnac2014}. Xie {\it et al} have reconstructed the 3D evolution of the time-integrated fluence distribution by controlling the relative position of the beam with the exit side of the sample \cite{XieSR2015}.

\subsubsection{Plasma luminescence}
Since the temperature of the plasma phase is typically several thousands Kelvins, blackbody emission is in the visible region of the spectrum. Side imaging of the plasma luminescence provides a semiquantitative characterization of the plasma distribution and temperature. This also allows for characterizing the dynamics in the multiple shot regime to follow the drilling mechanism \cite{Wu2002,Hwang2008}. 

Plasma emission can also include fluorescence from two-photon absorption \cite{Tzortzakis2006}, and fluorescence from the relaxation of self trapped excitons (STE's) and transient color centers (NBOHCs) \cite{Papazoglou2011}.

\subsubsection{Acoustic waves}
Direct recording of the amplitude of acoustic waves also provides indications on the laser deposited energy that eventually was converted to mechanical waves, and on the dynamics of the shockwave \cite{Noack1998,Kudryashov2008}.
Imaging of the dynamics of acoustic waves that follow shockwaves can be performed by shadowgraphy. The evolution of the wave speed in time provides estimations on the laser-deposited energy, using Sedov' theory. This is however restricted to very specific geometrical conditions.

\section{Microstructuring of transparent materials in multiple shot regime}
\label{sec:multishot}

In-volume structuring of transparent materials is requested for a number of applications where channels, deep trenches need to be processed: micro- or nano- fluidics, biosensors, fabrication of mechanical pieces with micron resolution, microelectronics, MEMS, micro-optics are typical fields of application.

When the typical transverse dimension exceeds 1~$\mu$m, multiple pulses are required for the structuration. In this regime, the propagation and absorption of a laser pulse is strongly affected by the modification of the material performed by the previous pulses. Indeed, the irradiation by an initial ultrashort laser pulse leaves index modifications, colored centers, highly absorbing defects, voids, ripples or a rough surface. On these structural defects, the next pulses can be scattered, absorbed, diffracted, guided, depending on the type of modification, and depending on numerical aperture and repetition rate.

Microstructuring with femtosecond lasers in multiple shot has been realized either from front or rear side surface of transparent materials. Front surface processing corresponds usually to ablation, drilling, trepanning or, for much weaker modifications, waveguide writing (see next chapter). This corresponds to a delicate geometrical configuration because of the reasons mentioned above. Scattering by structural defects can be extremely deleterious because ablation can happen even in regions that were not supposed to be illuminated.

\begin{figure}
    \centering
    \includegraphics[width=0.7\columnwidth]{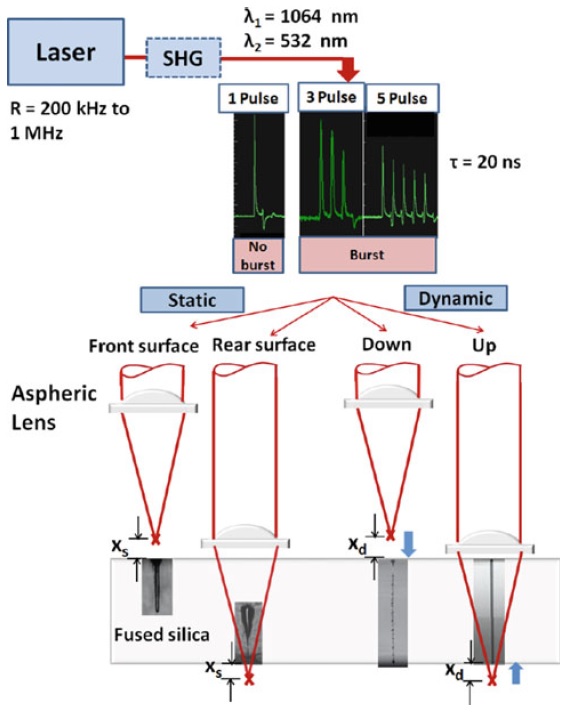}
    \caption{Concepts of the different laser drilling methodologies for high aspect ratio processing of transparent materials, involving static or dynamic sample positioning, and front or rear side processing. Reprinted figure with permission from \cite{Karimelahi2013} with courtesy of Prof. P. Herman. Copyright (2013) by Springer Nature.}
    \label{fig:multishot}
\end{figure}

In contrast, rear surface processing is very attractive because the problem of plasma plume shielding and the structural modifications induced by previous pulses have no impact on the propagation of the following pulses, except at the ablation site (see Figure \ref{fig:multishot}). The drilling strategy consists in illuminating the exit surface up to ablation and progressively translate the beam at the same speed as the one of debris removal. It has been successfully used to write high aspect ratio structures in glass or other transparent materials \cite{Zhao2011}.

Processing with high-repetition rate trains of pulses, {\it ie} bursts, takes advantage of the \index{heat accumulation} process \cite{Eaton2005} to increase the efficiency of the ablation or modification, so as to reduce the total amount of energy deposited in the material. This can reduce the size of the heat affected zone. Comprehensive comparison between front and rear surface processing with picosecond pulses demonstrated interesting processing parameter windows where high aspect ratio structures could be drilled with high speed with reduced heat affected zone. The performance on drilling was evaluated to be better with picosecond pulse durations in comparison with femtosecond ones, with comparable channel quality \cite{Karimelahi2013}.

When the aspect ratio is high, the energy density of the explosions at the ablation sites might be not enough to eject material out of the channel\index{water-assisted drilling}. To solve this issue, water assistance can help and remove debris out of the channels. This also allows for drilling non-straight channels in three dimensions \cite{Li2001,Kim2005}.  The aspect ratio of the channels drilled inside transparent solids can reach 1000:1 with diameters on the order of several microns, down to few hundreds of nanometers. In some configurations, the channel filled by water behaves as a resonance tube, and acoustic limitations are found when processing very high aspect ratio nanochannels. In this case, the length of the channel is limited by the node of the velocity inside the tube \cite{Lee2007}. As a remark, by using a high aspect ratio beam, such as a filament or a non-diffracting Bessel beam, the drilling can be performed without scanning. This technique is called self-guided drilling. This removes the constraint of translating the focal point at the same speed as the one of material removal \cite{Hwang2008,Bhuyan2010}.

A different strategy is based on a 2 steps process\index{etching}. The first consists in femtosecond laser modification of the fused silica matrix. Then, wet etching with hydrofluoric acid (HF) or potassium hydroxide (KOH) removes the modified glass \cite{Marcinkevicius2001}. This technique allows for creating various shapes in 3D, including high aspect ratio micro and nanochannels \cite{Bellouard2004,Ke2005,An2008,Wortmann2008}. Some groups have also taken advantage of both regimes, by laser processing the rear-side of the transparent workpiece set in contact with an etching liquid (HF or KOH) \cite{Cao2018}. However, the two-step approach is for now restricted to fused silica, several glasses and sapphire. The capability of the process relies on the difference in etching rates between laser-processed and pristine material.

In conclusion, multishot drilling regime is unavoidable for wide structures forming in transparent materials. Rear-side structuring removes some of the difficulties associated with the structural changes induced by previous pulses that are not easily predictable or controllable. For smaller scale structures, the situation is different and one can take benefit of generating voids in single shot regime.

\section{Single shot void formation under high numerical aperture focusing}
\label{sec:void}

In 1997, Glezer {\it et al} demonstrated for the first time that single pulses could generate nanometric cavities in the bulk of fused silica, quartz, BK7 glass, sapphire, diamond and acrylic \cite{Glezer1997}\index{void}\index{nanovoid}. 100 fs, 0.5~$\mu$J pulses were focused with a numerical aperture (NA) of 0.65, below the surface of the transparent materials. The diameter of the cavities was on the order of 200~nm, and the material surrounding the cavity was compressed. The authors suggested that extreme TPa pressures reached after laser energy deposition, could explain the formation of such voids in hard solid materials.
In silica glass, the voids could be moved and merged or be induced even with multiple shots \cite{Watanabe2000}.  

Besides the discovery of a new mechanism, these results were opening a new route for laser  processing of transparent materials. Indeed, they demonstrate the opportunity to process materials directly from the bulk instead of progressively processing, shot after shot, channels or long structures, from one of the surfaces.

\begin{figure}
    \centering
    \includegraphics[width=0.8\columnwidth]{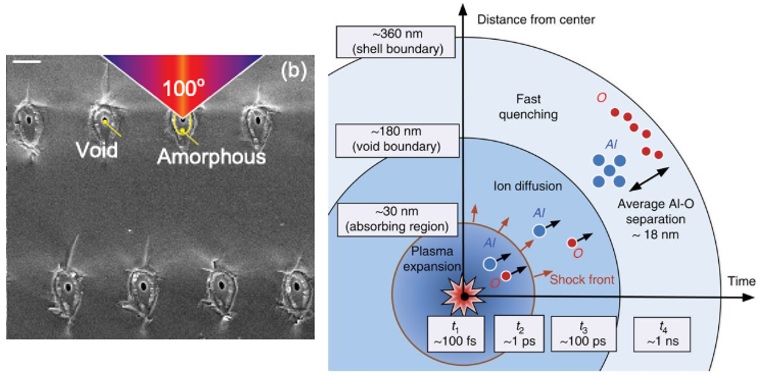}
    \caption{[left] SEM imaging of a pattern of nano-cavities created in sapphire by single 150 fs, 800 nm, 120 nJ pulses. The sample has been mechanically cleaved before imaging. Scale bar is 1~$\mu$m. Reprinted figure with permission from \cite{Juodkazis2006} with courtesy of Prof. S. Juodkazis. Copyright (2006) by the American Physical Society.[right] Concept of the micro-explosion, ion separation and ultrafast quenching developed by Vailionis {\it et al}. Reprinted figure with permission from \cite{Vailionis2011} with courtesy of Prof. A. Vailionis.}
    \label{fig:voids}
\end{figure}

Figure \ref{fig:voids}(left) shows an SEM image of such a nano-cavities produced in sapphire, visualized after mechanically cleaving the sample. A modified region can be observed around the spherical cavities. This modified region can be etched using an HF solution.  The formation of a cavity was interpreted in terms of shockwave release after the generation of a dense plasma, followed by a rarefaction wave. Figure \ref{fig:voids}(right) illustrates the concept. Hydrodynamic numerical simulations based on equation of states for fused silica \cite{Gamaly2006,Hallo2007} show that the formation of the void occurs in a typical timescale of a few hundreds of picoseconds after illumination. The shockwave\index{shockwave} stops when the internal pressure equals the Young modulus of the material. Separate experimental results  based on phase contrast microscopy characterizing the plasma dynamics were compatible with this theory \cite{Mermillod-Blondin2009}. 
Another potential formation mechanism is cavitation \index{cavitation}by material retraction under GPa pressure, in a similar way as what happens in water \cite{Vogel2008}.

In the model of nano-cavity formation after high-pressure shockwave and rarefaction wave, the pressures transiently reach teraPascals (TPa), and the compression of material around the void leaves a densified material. The density increase reach typically 14\% in sapphire \cite{Juodkazis2006}. This value was confirmed later in another geometrical configuration \cite{Rapp2016}.  The state corresponding to extreme pressures and temperatures is the Warm Dense Matter (WDM) state that lasts less than $\sim 1$~ns. The fast cooling can quench relaxation and can generate new material phases around the nano-cavity. Theoretical studies predict phase transitions of Aluminum into hcc-Al and bcc-Al at pressures in the multi-hundred GPa, confirmed recently by diamond anvil cell compression experiments \cite{Fiquet2018}. These phases of aluminium have been discovered around nano-cavities produced in Al$_2$O$_3$ \cite{Vailionis2011}, demonstrating compatibility of the high-pressure shockwave mechanism with experimental results.

In conclusion for this section, the formation of voids inside transparent materials reflects the potential for high energy density deposition within the bulk of transparent materials. A wide range of different structures are then possible provided that the propagation and energy deposition can be controlled. This is what will be discussed in the following sections.

\section{Filamentation and optical breakdown of Gaussian beams}
\label{sec:filamentation}

\index{filamentation}Filamentary structures, {\it i.e.} elongated damage tracks, were very early identified in the wake of high peak power laser illumination in dielectrics \cite{Hercher1964, Yablonovitch1972}. This was in fact a severe problem for ultrashort pulse amplification until the invention of Chirped Pulse Amplification \cite{Strickland1985}. During a long time however, optical breakdown and filamentation were opposed: optical breakdown regime was the mechanism where dielectrics undergo sufficient nonlinear ionization to induce a strong permanent modification \cite{Ashcom2006,Nguyen2003}. In contrast, filamentation was regarded as a dynamical mechanism that was transforming the initial Gaussian beam into a quasi-soliton. This regime was identified by a strong supercontinuum emission and low plasma density formation, such that the modifications generated are weak index changes, such as waveguides. 
\begin{figure}
    \centering
    \includegraphics[width =0.9\columnwidth]{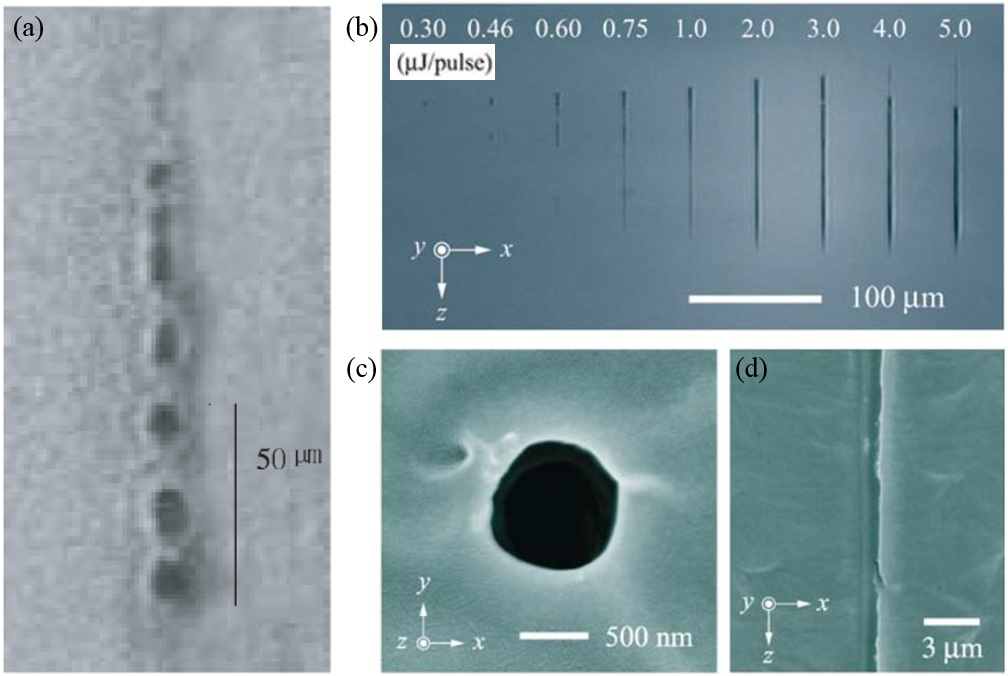}
    \caption{Diversity of damages produced in single shot by ultrashort pulses in the filamentation regime. (a)Non-periodic series of voids formed in fused silica. Reprinted figure with permission from \cite{Luo2001} with courtesy of Prof. Q. Gong. Copyright (2001) by the Institute Of Physics. (b-d) Void channel formation in PMMA. (b) Side view optical imaging of voids channels formed after single shot illumination for different input pulse energies. (c-d) Scanning Electron Microscopy (SEM) of the void formed for 2~$\mu$J: (c) transverse cross-section, (d) longitudinal cross-section. Reprinted figures (b-d) with permission from \cite{Sowa2005} with courtesy of Prof. W. Watanabe. Copyright (2005) by Springer Nature.}
    \label{fig:DiverseForms}
\end{figure}

However, filamentation has no precise definition \cite{Couairon2007}. Not only there is no precise boundary between optical breakdown\index{optical breakdown} and filamentation, but these regimes in fact do overlap \cite{Luo2001,Sowa2005}. It is specifically the regime of interest for this chapter. We can refer to filamentation as the regime of nonlinear pulse propagation in dielectrics, where dynamical reshaping of the pulse occurs in space and time under the influence of Kerr effects, nonlinear ionization and plasma defocusing, among others. 

Because Kerr effect\index{Kerr effect} in transparent dielectrics is three orders of magnitude higher than the one in air, and because plasma densities in solids easily reach 100 times the densities observed in air, filamentation in solids is much more confined than in air, and the filaments survive only some millimeters. While the diameter of plasma channels in gases is on the order of 100~$\mu$m, these are typically less than 10 ~$\mu$m in solid dielectrics \cite{Couairon2007}. Supercontinuum generation is not necessarily a condition for filamentation since this process is efficient only on very long propagation distances ($\sim$centimeters), when frequency mixing can become efficient.

In transparent dielectrics, a very wide family of modifications can be generated when the irradiation regime is complex.
Figure \ref{fig:DiverseForms} assembles typical results found in the literature. The morphology of these strong modifications (strong index changes, cavities) cannot be straightforwardly explained from the linear propagation of a Gaussian beam with which they have been produced. It is the filamentation that has reshaped the beam and induced these {\it a priori} unexpected morphologies. : elongated voids, channels,  series of voids (nonperiodic, periodic) \cite{Luo2001}.

\begin{figure}
        \centering
        \includegraphics[width=0.5\columnwidth]{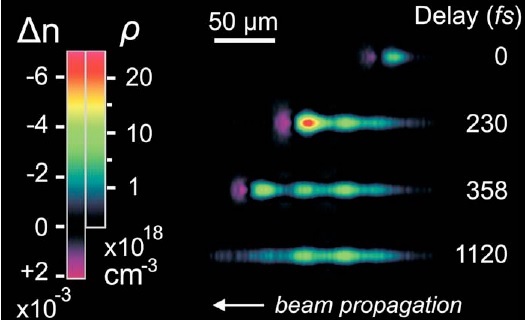}
        \caption{Holographic measurement of the spatial distribution of the plasma density at different pump-probe temporal intervals. Reprinted figure with permission from \cite{Papazoglou2008} with courtesy of Prof. P. Papazoglou and Prof. S. Tzorzakis. Copyright (2008) AIP Publishing.}
       \label{fig:Papazoglou}
    \end{figure}

The filamentation process can be understood from figure \ref{fig:Papazoglou}. The figure shows a measurement of the transient change of index of refraction in air during pulse propagation. We note that similar behavior can be observed in solids \cite{Papazoglou2014}. The positive index change is shown in purple, and it corresponds to Kerr self focusing at the rising edge of the pulse. Then, when the pulse intensity is sufficiently high, the medium is ionized and the plasma channel decreases the index of refraction. The negative index change tends to defocus the pulse, acting as a negative lens, so that the following part of the pulse generates slightly less plasma\index{plasma defocusing}. Because of the reduction of plasma generation, the defocusing effect is reduced and higher plasma density is generated. This process repeats itself as long as the intensity is sufficiently high \footnote{We note that this reasoning is somehow simplistic because it is based only on a spatial description, whereas in reality, the rising and trailing edges of the laser pulse do not experience the same effects.}. Depending on the exact spatial phase profile, the process of plasma generation might be quasi-periodic, very homogeneous or quite complex. It leaves a plasma channel which will relax in generating a modification of the material which morphology depends on the plasma density distribution. 

The competition between the different nonlinear effects that sustain the filamentation process can be evaluated with characteristic distances \cite{Couairon2007}. The nonlinear length is $L_{NL}=1/(n_2 k_0 I_0)$, where $n_2$ is the nonlinear Kerr index and $I_0$ the peak intensity of the pulse. The plasma characteristic length is $L_{plasma} = 2 n_0 \rho_c/(k_0\rho)$, 
where we use the notations of pages~\pageref{eq:epsilon_of_omega} and \pageref{eq:plasmaeq} and $\rho_c =\varepsilon _0 m_e \omega_0^2 /e^2 $ is the critical plasma density at the laser central frequency $\omega_0$.

 When these distances are on the same order of magnitude as the Rayleigh range inside the transparent material, then a rich dynamics can be induced. As an example for fused silica, for  peak intensities 10$^{12}$ to 10$^{13}$~W.cm$^{-2}$, the characteristic nonlinear length is on the order of some tens of microns, the plasma length shrinks from some 40~$\mu$m to some microns when the plasma densities increases from 10$^{19}$  to 10$^{20}$ cm$^{-3}$ as it is the case during the plasma buildup. Therefore, focusing with a numerical aperture below ~0.4 will trigger a long filamentation process, when spherical aberration is neglected. These number for instance match the experimental results of reference \cite{Papazoglou2014}. We note that most of the times, the Marburger formula \index{Marburger} for filamentation is mostly unapplicable \footnote{Marburger formula is calculated for a collimated beam. It does not take into account any spatial phase, like spherical aberration or focusing conditions. Dawes and Marburger formula is also semi-empirical \cite{Couairon2007} and therefore has a very narrow range of applicability.}.

Therefore, low focusing numerical apertures, short input pulse durations and high peak powers are prone to seed a filamentation regime with rich dynamics on long distance, where a number of four-wave mixing processes can take place, among others. 

\begin{figure}
    \centering
    \includegraphics[width=0.8\columnwidth]{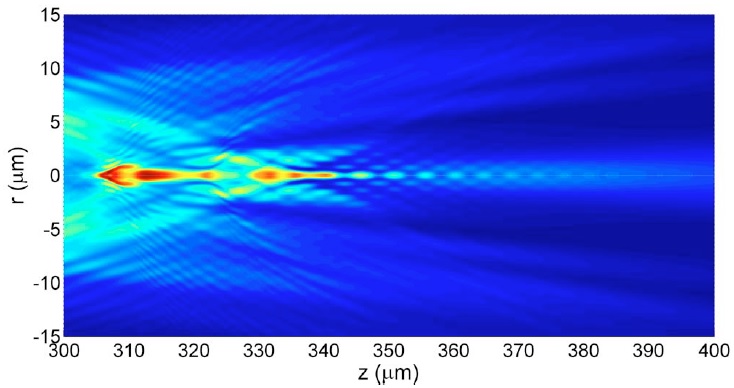}
    \caption{Numerical simulation of fluence distribution of focusing a Gaussian beam with NA 0.9 through 200~$\mu$m fused silica. Several high fluence spots appear along the optical axis. Reprinted figure with permission from \cite{Song2008} with courtesy of Prof. J. Qiu and Prof. Z. Xu. Copyright (2008) AIP Publishing.}
    \label{fig:PeriodicVoids_Song}
\end{figure}

Spherical aberration has an important contribution to trigger the filamentation process\index{spherical aberration}. This is particularly the case when a Gaussian beam is focused at the rear side of a thick sample. Under spherical aberration, paraxial rays are focused at a much farther point than the focal position of non-paraxial rays. This drastically elongates  the effective linear scale length. In turn, filamentation process can be triggered. As an example,  Ahmed {\it et al} inserted thick glass plates between the focusing microscope objective and the workpiece to induce long filaments in the glass workpiece \cite{Ahmed2013}. 
This is also the case for rear side focusing. In reference \cite{Luo2001}, the NA of 0.65 associated to rear side focusing formed a series of periodic voids (see Figure \ref{fig:PeriodicVoids_Song}). We note that it is with the same numerical aperture that Glezer et al used to generate single, well-confined, nano-voids \cite{Glezer1997}. Identically, Kanehira {\it et al} used NA 0.9 and focussing through 750~$\mu$m thick borosilicate glass and produced periodically spaced voids  \cite{Kanehira2005}.

Filamentation in transparent materials was demonstrated for a number of different laser wavelengths, ranging from IR to UV \cite{Tzortzakis2006}. The operable window is limited by the transparency window. Shorter wavelengths tend to generate more dense plasmas and reduce the filament length. A detailed study compares filament formation for IR and visible wavelengths \cite{Karimelahi2013}.

In the case of illumination with a pulse train, {\it ie} a burst, thermal effects play a role. Indeed, the typical cooling time after laser pulse irradiation is in the $\mu$s range (strongly depending on focussing conditions), such that the pulses within a burst of several MHz repetition rate regime influence themselves via thermo-optical effect. This effect increases the local index of refraction of glasses at the locations where the temperature is high \cite{Ghosh1995}. The heat accumulation\index{heat accumulation} can lead to laser beam trapping and guiding.

With a low repetition rate laser, the photo-excitation has completely relaxed before the arrival of the subsequent pulse. The latter diffracts on the structures left by the previous pulses. This regime was used to induce periodic damages \cite{Luo2001,Zhu2005,Kanehira2005,Sowa2005}. In this regard, we note that even a surface crater does not hamper occurrence of filamentation \cite{Luo2001}. 

In conclusion of this section, a filamentation dynamics is a complex phenomenon, highly dependent on input conditions and on the precise dynamics of ionization process. It is therefore extremely difficult to predict and to scale. Filamentation can generate plasma tracks with diverse morphologies. In the field of applications such as "filamentation cutting" or "filamentation welding", as we will see in section \ref{sec:applications}, state of the art usually refers to filamentation for the formation of long and homogeneous plasma channels on high aspect ratio.

Interestingly, the filamentation process, when it creates long uniform plasma channels, spontaneously transforms a Gaussian beam into a nonlinear conical wave \cite{Dubietis2004,Porras2008}. Nonlinear Bessel beams, characterized by a conical energy flow from the lateral rings to the central core, have been proposed as attractors to the filamentation regime \cite{Porras2004}. It is therefore natural to generate plasma filaments from Bessel beams, as we will describe in the next section.

\section{Nonlinear propagation of ultrafast Bessel beams}
\label{sec:Bessel}
Zeroth order Bessel beams\index{Bessel beam}\index{nondiffracting beams}\index{diffraction-free beams} are invariant solutions to the Helmholtz equation. Bessel beams can seed the nonlinear propagation regime where a homogeneous plasma channel is generated \cite{Courvoisier2016,Duocastella2012}. In this section, we will review what are Bessel beams, highlight the properties of their propagation in the nonlinear regime which are the most relevant for laser materials processing.  Then we will review basic applications, particularly high aspect ratio nanochannel processing.

\begin{figure}
    \centering
    \includegraphics[width= 0.9\columnwidth]{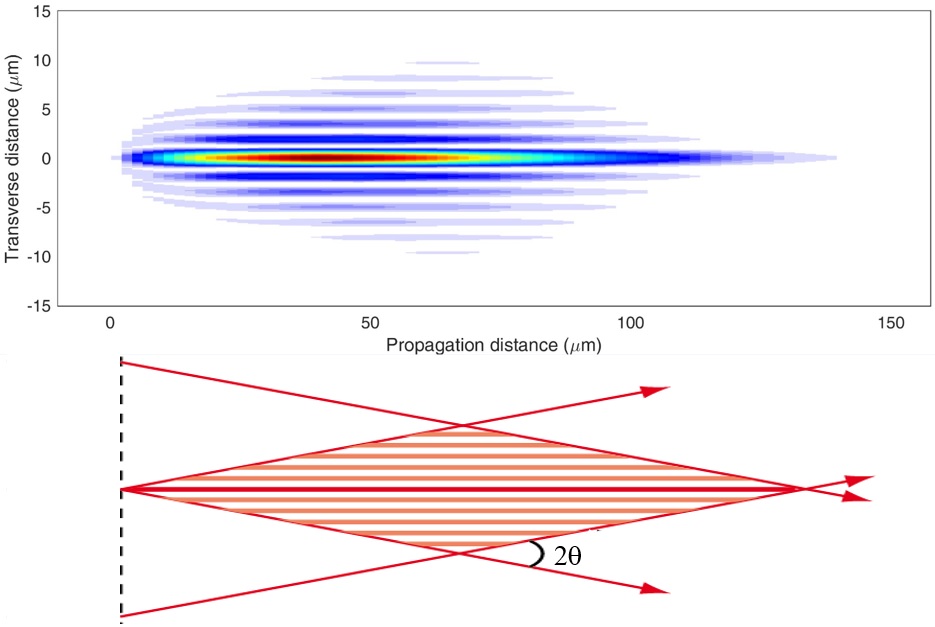}
    \caption{ (top) Intensity distribution of a Bessel-Gauss beam (bottom) corresponding ray-tracing representation, showing that the Bessel beam is an interference field with cylindrical symmetry.}
    \label{fig:Bessel_Interference}
\end{figure}

\subsection{Bessel beam structure}

Within a scalar model of monochromatic light, Durnin demonstrated that the Helmholtz equation $\left(\nabla ^2+ (\omega/c)^2 \right) A=0$, has a solution that is propagation-invariant with a hot spot. This central hot spot that can be of diameter down to "a few wavelengths", as he wrote, but in fact even below the wavelength \cite{Durnin1987a,Durnin1987}. The solution found by Durnin is cylindrically symmetric: 

\begin{equation}
A(r,z)=J_0(k_0 \sin \theta r) e^{i k_0 \cos \theta z}
\end{equation}

\noindent where $k_0$ is the wavevector and $\theta$ is the Bessel beam parameter, which is called the {\it cone angle}.

This solution, as it is the case for plane waves, is of infinite energy. We can experimentally generate only the apodized solutions. Several types of apodizations exist, which depend on the mean of generating the finite energy Bessel beam. In the rest of this chapter, finite energy Bessel beams will be referred to as "Bessel beams" for sake of simplicity.

The first experimental realization of Bessel beam was from an axicon \cite{McLeod1954}\index{axicon}, even before it was realized that this was corresponding to a "diffraction-free" solution. Durnin {\it et al} produced Bessel beams from a ring slit, placed at the focal plane of a converging lens which Fourier transformed the ring aperture into a Bessel beam. Indeed, in the spatial frequencies ($k_r$) space, {\it i.e.} the Fourier space, an ideal Bessel beam is a circle of amplitude $A(k_r)=\delta(k_r-k_0\sin\theta)$. Because of the properties of the Fourier transform, the thinner is the ring slit, the longer is the actual Bessel beam length. However, this mean of generation has poor energy throughput since most of the power is lost. In the field of laser materials processing, it is preferable to shape beams in the {\it direct} space, by opposition to the Fourier space.

Bessel beam generation from the direct space can be performed using axicons \cite{Grunwald2004,Grosjean2007,Tsampoula2007,Akturk2009,Xie2012,Duocastella2012}, holograms \cite{Vasara1989}, or using Spatial Light Modulators \cite{Chattrapiban2003, Courvoisier2009} or, equivalently, Diffractive Optical Elements (DOEs)\cite{Amako2003,Khonina2004}. The shaping technique consists in applying a spatial phase $\phi(r) = k_0 r \sin \theta$.
The application of such phase onto a Gaussian beam creates what is called a Bessel-Gauss beam\index{Bessel-Gauss beam}. The evolution of the intensity as a function of the propagation distance $z$ can be derived on the optical axis from the stationary phase approximation of Fresnel diffraction integral:  
\begin{equation}
I(r=0,z)=4 P_0 k_0 z \sin^2\theta e^{-2(z \sin\theta/w_0)^2}/w_0^2  
\end{equation}

\noindent where $P_0$ and $w_0$ are respectively the power and the waist of the input Gaussian beam \cite{Roy1980,Jarutis2000}. High quality axicons enable the generation of high-power Bessel-Gauss beams without spatial filtering \cite{Boucher2018}.

Figure \ref{fig:Bessel_Interference} shows a ray tracing representation of a Bessel-Gauss beam. A Bessel beam is an interference field, which longitudinal extent, the {\it Bessel zone}, is: $Z_{max}\sim w_0/\tan\theta$. It is apparent from this geometrical concept that the invariance of the waves crossing angle along the propagation, makes the fringes period invariant. In other words, the central spot size does not change along the interference field, hence the denomination "diffraction-free". In contrast with Gaussian beams, Bessel-Gauss beams have two free parameters to independently adjust the Bessel zone length and the diameter of the central spot. The latter is $d_{FWHM} \sim 0.36 \lambda_0/\sin \theta$ only determined by the cone angle, whereas the Bessel beam length can be independently adjusted by the input beam waist.
It is important to realize that a Bessel beam corresponds to a line focus. Each point on the focused segment is topologically linked to a circle in the input plane \cite{Froehly2014}. In this regard, the energy does not flow along the optical axis, but instead the energy flow is conical.

The polarization state of a Bessel beam is close to the one of the input beam, since, with sub-30$^{\circ}$ cone angle, the propagation is close to paraxial \cite{Zhang2007}. The longitudinal component of the electric field is mostly negligible in the experiments described below.

We note that upon refraction from air to a dielectric medium, both the wavelength and cone angle are corrected by the index of refraction of the dielectric. But these corrections cancel out and do not change the value of central spot size in the material \cite{Brandao2017}. In contrast, the length of the Bessel zone is increased by the factor $n_0$.  This is similar to the case of a Gaussian beam where the Rayleigh range is increased while the waist remains identical upon refraction \cite{Nemoto1988}.

Up to here, we have described monochromatic Bessel beams. In principle, the bandwidth of ultrashort pulses has to be taken into account in the description. But since it is  less than 1\% for pulses of 100~fs, spatio-temporal aspects of the pulse generation can be neglected. Apart from the fact that the on-axis wavepacket created by Bessel-X-Pulses and Pulsed Bessel beams does not travel at the same speed (respectively $c/\cos \theta$ and $c \cos\theta$), no impact in terms of plasma generation efficiency has been reported up to here. More details can be found in references \cite{Klewitz1998,Froehly2014}.

\subsection{Filamentation of Bessel beams}

The nonlinear propagation of Bessel beams can be described in terms of 3 different families \cite{Polesana2008}. {\it Weakly nonlinear Bessel beams} generate negligible plasma density and are only characterized by central lobe shrinking due to Kerr effect; {\it nonstationary Bessel beams}, as for their denomination by Polesana {\it et al}, generate a quasiperiodic plasma distribution along the propagation in the material. The third family, {\it stationary Bessel beams} is characterized by a quasi-invariant propagation, that generates a homogeneous plasma channel.

In more detail, the second regime is largely driven by Kerr nonlinearity which generates via four wave mixing processes, two secondary waves with transverse wavevectors $k_r = \sqrt{2} k_{r0}$ and $k_r = 0$, where $k_{r0}=k_0 \sin \theta$ is the radial wavevector of the initial Bessel beam \cite{Gaizauskas2006,Ouadghiri-Idrissi2017}.  The interference between these secondary waves and the initial Bessel beam create periodic oscillations \cite{Polesana2008, Norfolk2010}. Periodic modifications in glass have been demonstrated in this regime by Gaizauskas et al \cite{Gaizauskas2006}.

The third regime is the most interesting for micro-fabrication. It is indeed the regime where enough losses occur in the central lobe to stabilize the dynamics. A conical flow of energy, oriented from the lateral lobes to the central one, compensates the energy loss.  This regime corresponds to the monochromatic Nonlinear Unbalanced Bessel Beams (NLUBB). Indeed, a Bessel beam can be seen as a superposition of two cylindrical Hankel waves with equal weights, one propagating inward and the other propagating outward \cite{Porras2004}. In a NLUBB, energy loss within the central lobe reduces the weighting coefficient for the outward component. This reduces the contrast of the fringes and implies a net energy flow towards the center. Noticeably, in this regime, the spatio-temporal shape remains quasi-invariant all along the propagation, hence the denomination {\it stationary Bessel beam}. This is in contrast with the second regime, {\it non-stationary Bessel beam}, where a periodic reshaping strongly modifies the spatial temporal shape of the pulse \cite{Polesana2008}.

The propagation-invariant NLUBB solution was proposed as an attractor to the filamentation regime \cite{Porras2004}. This solution cannot be found for all input parameters, but the operating window, in terms of peak power, is wider when the cone angle is increased.  Indeed, for higher peak powers, nonlinear losses are higher which tends to reduce the impact of Kerr nonlinear dynamics. This has an important impact for applications to laser materials processing: high powers are needed to generate high plasma densities. A stationary regime can be reached for a given high peak power if the Bessel cone angle is sufficiently large.

\subsection{High aspect ratio processing with propagation-invariant ultrafast Bessel beams}

The stationary filamentation regime was early recognized to have strong potential applications \cite{Porras2004}. Early works with ultrafast Bessel beams in glass - before the theoretical work on Bessel filamentation - have shown that it is possible to write index modifications without the need for translating the sample \cite{Marcinkevicius2001a,Amako2005}.

With much higher cone angles, we demonstrated the ability to create high aspect ratio nanochannels\index{nanochannel} with a single femtosecond pulse \cite{Bhuyan2010}. In borosilicate glass, the channels could be drilled either from an entrance or exit surface, with diameters ranging from $\sim$200~nm to 800~nm. The diameter could be tuned quasi-linearly with input pulse energy. The aspect ratio reached at that time 100:1, which was in line with the aspect ratio of the beam. Through channels could be drilled also with a single pulse, and periodic arrangements of nanochannels could be realized.


\begin{figure}
    \centering
    \includegraphics[width=\columnwidth]{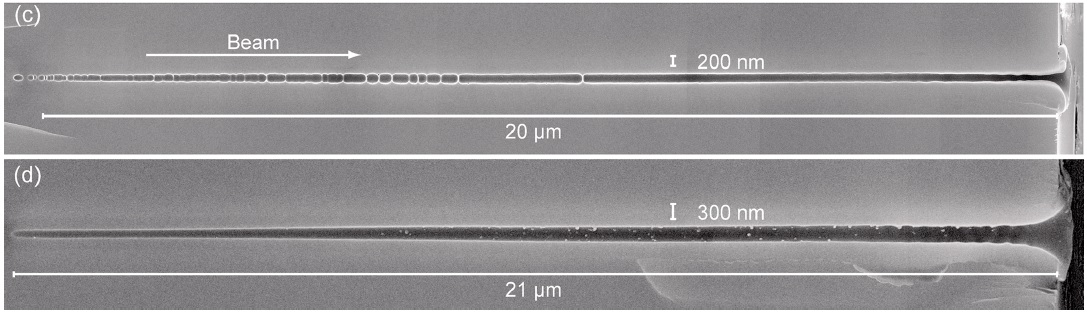}
    \caption{ From . Open channels formed after femtosecond illumination ($\sim$230~fs) by Bessel beams with cone angle $\theta_{\texttt{glass}} = 17^{\circ}$ in Corning 0211 glass, for two pulse energies. SEM imaging is performed after mechanical cleaving. Reprinted figure with permission from \cite{Bhuyan2010}. Copyright (2010) AIP Publishing }
    \label{fig:ChannelsBessel}
\end{figure}

In borosilicate glass, it was possible to generate a channel only if the Bessel beam was crossing one of the sample surfaces, {\it i.e.} only if the channel was opened on one of the sides. In contrast, in sapphire, it was possible to create a high aspect ratio nano-void  fully enclosed within the materials bulk. In this case, the void is formed only by compression of the material around. Yet the void formation process in this configuration is not yet fully understood, we infer that the 10- fold higher thermal diffusion coefficient of sapphire allows for fast cooling. This can prevent cavity closing, in contrast with the case of borosilicate glass.

\begin{figure}
    \centering
    \includegraphics[width=\columnwidth]{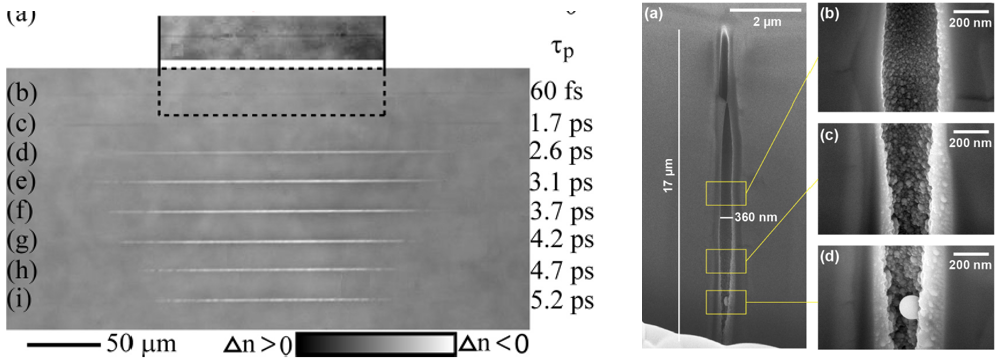}
    \caption{(left) Phase contrast images of high aspect ratio structures formed after illumination by Bessel beams with cone angle $\theta_{\texttt{glass}} = 8^{\circ}$ in 7980-5 F Corning glass, for different pulse durations. Note the large difference in cone angle with reference \cite{Bhuyan2010}. Reprinted figure with permission from \cite{Bhuyan2014} with courtesy of Dr. R. Stoian. Copyright (2014) AIP Publishing. (right)  High aspect ratio void formed in sapphire after illumination by a Bessel beam of cone angle $\theta_{\texttt{sapphire}} = 15^{\circ}$ and pulse duration 3~ps. The heat affected zone is far more pronounced than in the femtosecond case and the sides of the channel evidence the formation of phase transformations during the cavity formation process in this case. From \cite{Rapp2016}, Creative Commons licence.}
    \label{fig:ChannelsPico}
\end{figure}

Further investigations with picosecond pulses have been independently performed by several groups in a number of different glass materials. Interestingly, it seems that picosecond pulses generate channels that are more visible under optical microscopy than the ones created by shorter femtosecond pulses (see Figure \ref{fig:ChannelsPico}(left)). A parametric study of the channel morphology as a function of input pulse energy and pulse duration has been reported in reference \cite{Bhuyan2014}. The pulse duration was adjusted by temporally stretching a femtosecond pulse and the Bessel beam aspect ratio was $\sim$1000:1.  It was found that in this case, multi-picosecond pulse durations could create uniform voids. For too high pulse energies, fragmentation of voids was observed. For very short pulses, less than 100~fs, the formation of empty channels was less clear. In this case, we stress that characterization techniques are at limit of resolution.

In parallel, nanovoids induced by 3~ps pulses in sapphire\index{sapphire} were characterized by FIB milling process. The result is shown in figure \ref{fig:ChannelsPico}(right). It is apparent that the morphology of the cavity is highly different from the case of femtosecond pulse illumination. Nanoparticles accumulated on the walls of the cavity are clearly observable, as well as a very wide heat affected zone \cite{Rapp2016}.
It is too early to determine if the more apparent damages produced by picosecond pulses with respect to femtosecond ones, arise from a different energy density deposition and/or from a different photo-excitation pathway.

Experimental time resolved phase contrast microscopy opened new perspectives on the formation of the void. Bhuyan {\it et al} imaged the transient index distribution at times case ranging from nanoseconds to microseconds \cite{Bhuyan2017}. They conclude that the void opening is slow in comparison with the shockwave theory. They infer that the void formation in this 2D case arises from cavitation of a low viscosity liquid phase. The main difference in comparison with the shockwave\index{shockwave} theory is that the estimated deposited energy density is of $\sim$7~kJ.cm$^{-3}$, which is in high contrast with the values estimated in the case of spherical void formation, on the order of 90~kJ.cm$^{-3}$ \cite{Hallo2007}.

Wang {\it et al} have investigated by shadowgraphy the mechanical wave ejected after the plasma formation in PMMA. They observed a wave with speed corresponding to the speed of sound in PMMA, whatever the input pulse energy. This is compatible with both theories on cavity formation since in the shockwave case,  the latter is supposed to propagate only over less than a few microns, {\it i.e.} below the resolution of the shadowgraphy experiment \cite{Wang2017}.

\section{Applications}
\label{sec:applications}
\index{filamentation}
\index{Bessel beam}
Single shot generation of plasma columns of $\sim$1$\mu$m diameter and length several tens to hundreds of micrometers has a number of different applications that we will review here. As mentioned earlier, the plasma channel generated by a smooth regime of filamentation from a Gaussian beam is quite close to the one generated by a Bessel ultrafast pulse. As we have seen above, the difference lies in the ability to control independently the parameters (length, diameter, pulse duration) that makes the Bessel beams attractive. We will treat the applications of both types of filaments in a single section. Most of the applications were started with Gaussian filaments and refined more recently with Bessel or Bessel-like beams.
\subsection{High aspect ratio refractive index modifications}

At relatively low peak power, long plasma channels have been used to write a series of index modifications in glass and polymers. The process is applicable in most of the transparent materials. However, as for Gaussian beam focusing, the positive or negative sign of the photo-induced refractive index change depends on the material itself.\index{grating}

\begin{figure}
    \centering
    \includegraphics[width = \columnwidth]{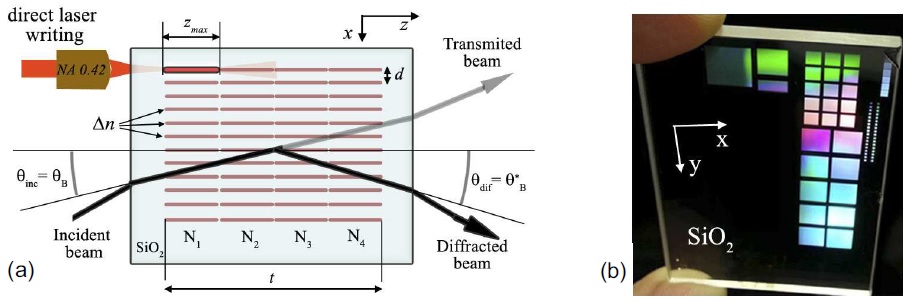}
    \caption{(left) Concept of the Bragg grating writing by a Bessel-Gauss beam. Several layers form a thick grating. (right)  Optical view of gratings written in fused silica with different parameters.Reprinted figure with permission from \cite{Mikutis2013} with courtesy of M. Mikutis, Workshop of Photonics. Copyright (2013) by the Optical Society of America.}
    \label{fig:grating}
\end{figure}

Long plasmas tracks have been used for instance to fabricate gratings in a number of different materials: PMMA \index{PMMA}, silica glass \index{glass}, or even chalcogenides\index{chalcogenides} (see Figure \ref{fig:grating}) \cite{Mikutis2013,Matushiro2016,Zhang2018} where the empty channels formed by Bessel beam illumination were used as scatterers in the vicinity of a waveguide \cite{Martin2017}. 

\subsection{Ultrafast laser welding}
\index{welding}
Joining transparent materials such as two types of glasses, or joining a transparent material on silicon or metal, is a need for a very large number of applications fields: opto-fluidics, biological analysis, microelectronics, MEMS, MOEMS require that structured glasses or silicon and metals have to be sold together after the microstructuration. Despite a number of different joining techniques exist, none allows for joining over only a few micrometers width. 
Ultrashort pulse lasers are ideal tools for this application, because they can melt the transparent material with very high spatial resolution, while preserving the optical, mechanical, electrical properties of the surrounding components.  Before welding, the two parts to be welded have to be set in tight contact. Then, laser illumination is used to melt the transparent material which expands and fills the empty space. After cooling, which scale is in microseconds, the two pieces are welded together.

\begin{figure}
    \centering
    \includegraphics[width=\columnwidth]{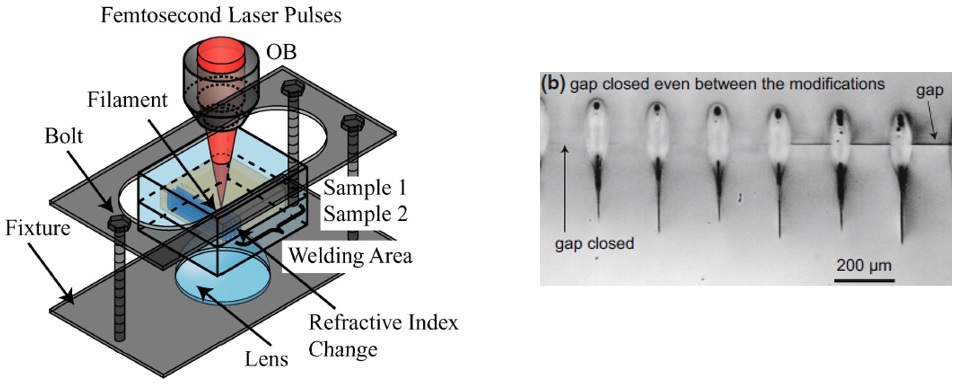}
    \caption{(Left) From  Concept of ultrafast laser welding of glass with filamentation. Reprinted figure with permission from \cite{Tamaki2006} with courtesy of Prof. W. Watanabe. Copyright (2006) by the Optical Society of America.(Right) From . Example of side view imaging of welded glasses. Depending on the melted pool position, the melt glass could fill the gap even between irradiation sites. Reprinted figure with permission from \cite{Richter2015} with courtesy of Prof. S. Nolte. Copyright (2015) by Springer Nature.}
    \label{fig:welding}
\end{figure}

The filamentation welding technique benefits of the relatively high volume of heated material in the plasma column, together with the relaxation of the positioning constraint \cite{Tamaki2005}, as shown in Figure \ref{fig:welding}(left). Dissimilar materials have been welded \cite{Watanabe2006}, even glass on silicon or metals \cite{Tamaki2006}. Welding with gaps up to $\sim$3~$\mu$m has been successfully achieved using bursts and heat accumulation effect \cite{Richter2015}, as shown in Figure \ref{fig:welding}(right).

Mechanical characterizations demonstrate that this technique is extremely powerful because, in terms of traction, the weld parts can be as strong as the bulk material itself.  The strength of the weld depends on the difference of the thermal and mechanical properties of the two materials. Large differences obviously have a negative impact on the strengths of the bonding\index{bonding}. We note that the use of burst and heat accumulation\index{heat accumulation} effect tends to relax the stresses left in the material and provide stronger welding \cite{Richter2015}.

\subsection{Stealth dicing of brittle transparent materials}
\index{stealth dicing}
\index{glass cutting}
\index{cutting}
High speed separation of materials is a key important technology for a number of applications, specifically for mass fabrication such as screen covers, touchscreens, electronics or lightning technologies. A specific need is to separate at high speed glass sheets with thickness of several hundreds of micrometers. In order to preserve the resistance of glass to bending and other types of stresses (thermal shocks), the cut needs to be free of chipping, with limited defects in the vicinity of the cut surface. 

"Stealth-dicing" is a technology initially developed for high speed, ablation-free silicon wafer cutting for the microelectronics industry \cite{Kumagai2007}. The concept is that a laser, which wavelength is chosen in the transparency window of the material ({\it i.e.} IR laser for silicon), generates a plane of defects within the depth of the material. When the material is set under thermal or mechanical stress, it cleaves along this plane. The initial technology was based on nanosecond IR lasers, and the morphological damages in silicon were extending typically on the scale of tens of micrometers.

\begin{figure}
    \centering
    \includegraphics[width =\columnwidth]{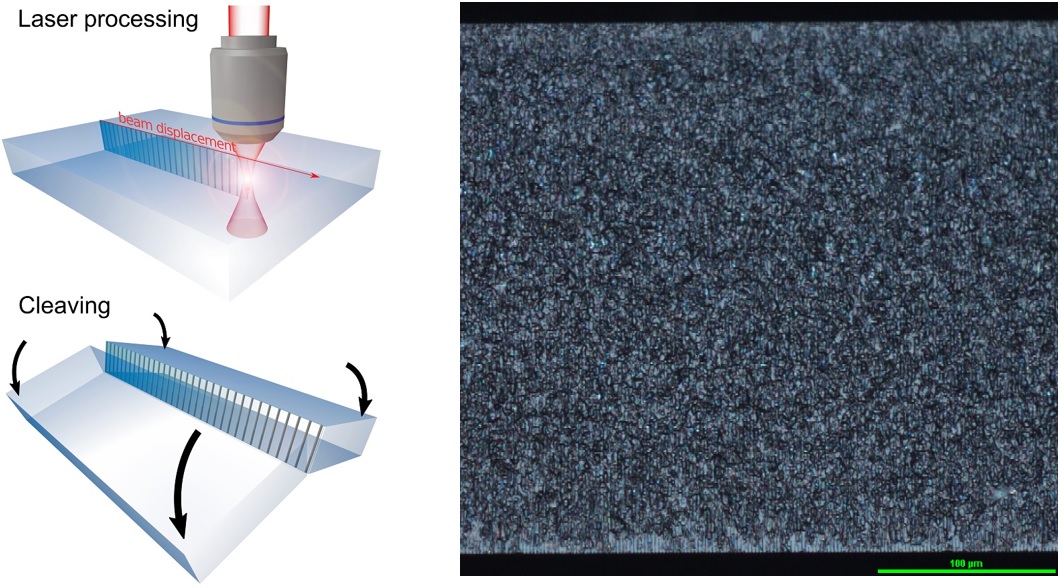}
    \caption{(left)  Concept of stealth dicing : in a first step, high speed laser processing creates a series of nanochannels aligned in a plane, which guides cleaving under mechanical stress. Courtesy of R. Meyer, {\it FEMTO-ST Institute}, France. (right) Example of optical microscopy view of gleaved glass after processing. With courtesy of J. Safioui, {\it FEMTO-Engineering}, France }
    \label{fig:StealthDicing}
\end{figure}

A similar technique was developed to separate glass, based on filamentation and plasma channel formation, leaving high aspect ratio nanovoids in glass. A periodic pattern of voids, separated by $\sim$5 to $\sim$25~$\mu$m, allows the material to be mechanically cleaved. This can be performed also with Bessel beams \cite{Bhuyan2015,Mishchik2017}.  Using commercial ultrafast lasers with multi-100 kHz repetition rate, it is feasible to irradiate at a speed on the order or exceeding 1m.s$^{-1}$. A small mechanical stress is enough to separate glass pieces, as shown in figure \ref{fig:StealthDicing}. 
This technology is particularly attractive in the case of chemically strengthened glass, such as the glasses used for cover-screens of smartphones, since the glass self-cleaves after laser processing.
 After cleaving, the walls are straight and free from chipping. The technique is mostly non-ablative, which avoids the issues of cleaning debris. Noticeably, it is also possible to cleave along curved paths. 

 To shape the processed glass or cut with angles different from 90$^{\circ}$,  illumination at non-perpendicular direction is desirable. But in this case, the non-uniform optical path difference over the beam cross-section  restricts the length of a Bessel beam inside the transparent workpiece. Jenne {\it et al} have developed an approach where the optical phase profile of an initial Bessel beam is compensated by a secondary mask. Cut with tilted angles were demonstrated up to 30 degrees \cite{Jenne2018}.

At high input pulse energies, the energy  stored in the material is sufficient to generate cracks.  A slight asymmetry in the input beam is sufficient to make the crack direction deterministic instead of random. This property has been exploited by Dudutis {\it et al}, by using an imperfect axicon\index{axicon}, which generates a non-circularly symmetric Bessel beam. It was used to generate cracks extending transversely up to 100~$\mu$m away from the central nanochannel \cite{Dudutis2016}. This brings the potential to increase the inter-channel distance for stealth dicing of glass at even higher speeds. Heat accumulation using burst mode with Bessel beams was also used to initiate the cracks \index{cracks} \cite{Mishchik2017}.

\begin{figure}
    \centering
    \includegraphics[width = 0.9\columnwidth]{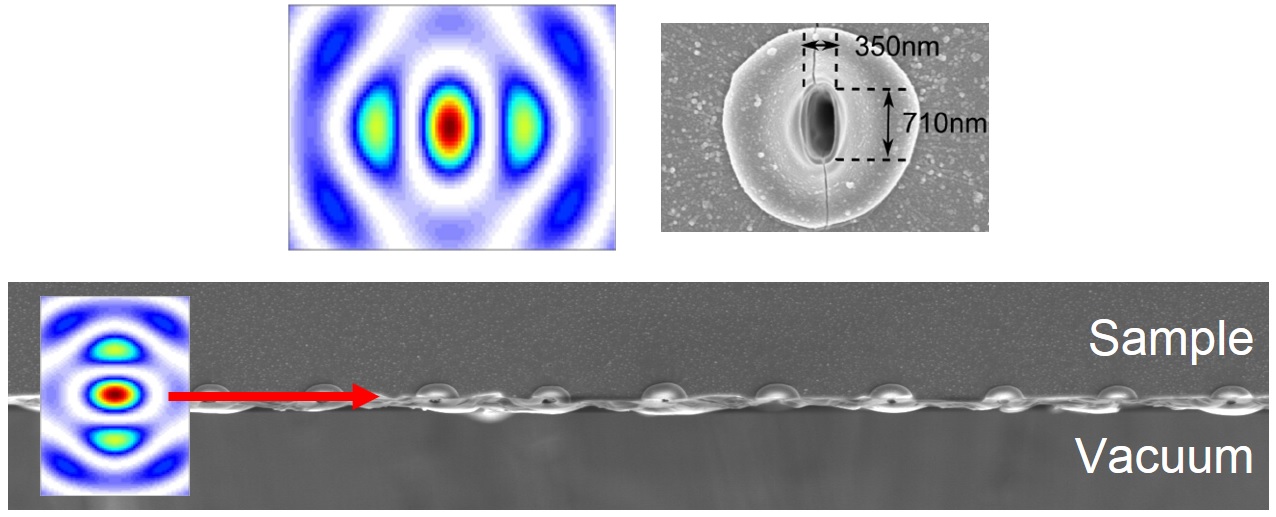}
    \caption{(top row) Transverse cross-section of an elliptical Bessel beams and corresponding SEM image of elliptical channel produced by the beam with single pulse illumination in glass. (bottom) The beam and red arrow show the laser scanning configuration. SEM image : top-view of a glass sample cleaved by stealthdicing technique, where it is apparent that all elliptical channels were processed.  With courtesy of R. Meyer, {\it FEMTO-ST Institute}, France.}
    \label{fig:EllipticalBessel}
\end{figure}

Instead of using crack formation guided by an imperfection in the axicon, it is also possible to create an asymmetry in the Bessel beam, using spatial filtering, so that the generated non-diffracting beam has an elliptical cross section. Using $\sim$3~ps single pulse illumination, such beams generate nanochannels in glass also with elliptical cross-sections, whose ratio major/minor axis is the same as the beam ratio \cite{Meyer2017}. The elliptical cross section allows for enhancing the mechanical stress at the tips of the ellipses and increases the reliability of stealth dicing. A detailed statistical study also demonstrated that the cleaving was requiring less mechanical deformation in this case, with the second benefit of leaving less defects in the processed glass, since all laser-induced channels are perfectly cleaved through \cite{Meyer2017a}, see figure \ref{fig:EllipticalBessel}.
\index{Bessel beam, elliptical}

\subsection{Separation of sapphire}
\index{sapphire}
Sapphire is an important technological material. Its high hardness, just below the one of diamond, make it an ideal cover for screens or for watches. This crystal is even more importantly used as a substrate for the growth of LEDs. Sapphire was also processed with the same stealth dicing technique as described in the previous sub-section.  A complementary approach is to take benefit of the crystalline structure of sapphire to guide the fractures. As in the case of glass, laser illumination with high pulse energies generate cracks even in the single shot regime. For C-cut sapphire, 3 crack directions are usually observed with Bessel beam illumination along the c-axis. However, below a pulse duration of $\sim$600~fs, the fracture can occur in a single direction, jointly determined by the laser pulse polarization and the scanning direction. This was exploited to initiate a series of cracks with very large inter-pulse distance (25 ~$\mu$m) paving the way for higher speed cutting \cite{Rapp2017}.\index{cracks}


\subsection{Structuration of diamond}
\index{diamond}
\index{graphitization}
\index{electrode}
Diamond is the hardest material and it is extremely difficult to process. It has a number of applications, particularly because it is bio-compatible. It is also increasingly used in quantum photonics . Ablation of diamond is for now still performed from the surface \cite{Kumar2018}, no high aspect ratio void formation has been yet reported to the best of the author's knowledge.

Diamond has been also proposed as a new material to build high-energy particle detectors. For this application, conductive graphite wires are needed in the bulk of the material. Graphitization of the bulk material has been successfully achieved with ultrafast Bessel beams. A single, 10~$\mu$J pulse was sufficient to create a conductive column through 500~$\mu$m diamond sample \cite{Canfield2017}. We remark that surface and bulk graphitization is a phenomenon that builds up from pulse to pulse as described in reference \cite{Kumar2017}.

\begin{figure}[htb]
    \centering
    \includegraphics[width =\columnwidth]{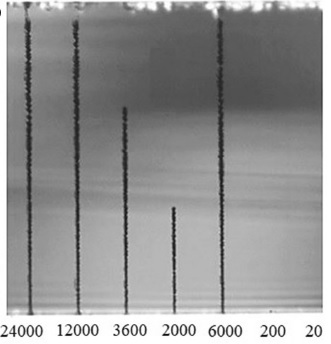}
    \caption{From . Optical microscopy views of graphitization marks created in the bulk of 500~$\mu$m thick diamond, with Bessel pulses of energy 3.5~$\mu$J. The number of pulses at 20~Hz repetition rate is indicated below each graphitized column. Reprinted figure with permission from \cite{Kumar2017} with courtesy of Dr. O. Jedrkiewicz. Copyright (2017) by Springer Nature.}
    \label{fig:diamondGraphitization}
\end{figure}

\subsection{Processing of silicon}
\index{silicon}
Silicon is a material of major interest for microelectronics and has a immense field of applications. Specifically, there are needs in the fields of creating waveguides for silicon photonics as well as micro and nanochannels for the cooling of silicon chips or to insert conducting electrodes transmitting signals from one side to the other.  These Through Silicon Vias (TSV)\index{Through Silicon Via (TSV)}\index{TSV, Through Silicon via} are particularly important for next generation 3D microelectronic chips.

Silicon is transparent  in the infrared region of the spectrum, for wavelengths higher than $\sim$1.1~$\mu$m. In this context, attempts on reproducing the results obtained in dielectrics were performed with femtosecond Bessel beams with a central wavelength of 1.3 ~$\mu$m. However, an absence of morphological modification was observed for bulk focussing with ultrafast pulses. This was explained by the authors as originating from the strong two-photon absorption \cite{Grojo2015}. Recently, Tokel {\it et al}, processed modifications in 3D, opening routes to similar processing as in glass, but this was with nanosecond pulse durations and requires a nonlinear feedback mechanism involving the rear surface of silicon \cite{Tokel2017}.

\begin{figure}
    \centering
    \includegraphics[width = \columnwidth]{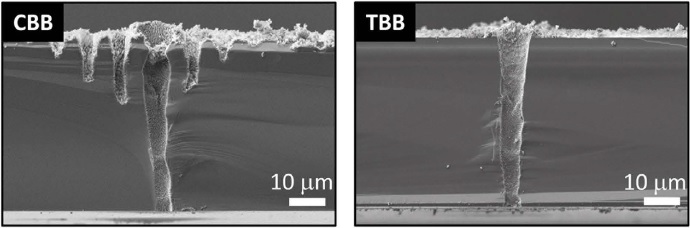}
    \caption{Silicon processing with Bessel beams in the multishot regime. (left) SEM view of a Through Silicon via (TSV) in silicon processed with a conventional Bessel beam (CBB). (right) same view for a Tailored Bessel beam (TBB) where the lobes of the Bessel beam have been removed. Reprinted figure with permission from \cite{He2017} with courtesy of Prof. K. Sugioka and Prof. Y. Cheng. Copyright (2017) Creative Commons licence.}
    \label{fig:silicon_Bessel}
\end{figure}

Bessel beams were also investigated for TSV drilling for a laser central wavelength of 1.5~$\mu$m (Figure \ref{fig:silicon_Bessel}. As the drilling with conventional Bessel beam was showing not enough contrast between the  lobes, an apodized version of Bessel beam was developed and 10~$\mu$m diameter TSV in 100~$\mu$m thick silicon wafer was drilled with $\sim$1200 laser pulses at a repetition rate of 1~kHz \cite{He2017}. 

More recently, bulk modifications in more conventional Gaussian-beam approach were demonstrated based on three different processes.  In reference \cite{Chanal2017}, illumination by a numerical aperture close to 3 could induce an index change in the bulk of silicon with a single pulse. In the multiple shot regime, 250~kHz repetition rate illumination with 350~fs pulses enabled to produce waveguides in silicon \cite{Pavlov2017}. The buildup of index modification was shown to be more reliable with 10~ps pulses instead of shorter pulses \cite{Kaemmer2018}. The mechanisms leading to modification of the silicon bulk are still incomplete and more experiments are needed to provide a clear overview for bulk laser processing of silicon.

\newpage
\section*{Conclusion}
In conclusion, the extremely high peak power of ultrashort laser pulses makes it possible to deposit energy with 3D control inside the bulk of transparent materials. This can be used to generate waveguides, nanogratings or even nano-cavities. Ultrashort laser pulses are therefore well-suited to answer the needs for high aspect ratio micro- and nano- processing, for drilling, cutting, producing channels for micro-nano fluidics or microelectronics. We have reviewed the basic mechanisms of pulse propagation and plasma formation inside transparent materials, as well as the experimental characterizations. For wide structures, high aspect ratio laser processing requires the multiple shot illuminations regime. The best condition corresponds generally to process from the exit surface of the workpiece, potentially with assistance of a liquid or an etchent.

Breakthroughs in the field have been made with single shot or single burst processing with filamented beams, which create long and thin homogeneous plasma channels. The structures that are generated therefore possess a very high aspect ratio. Predictable filamentation is made possible with "nondiffracting" Bessel and Bessel-like beams. Control on single shot filamentation has enabled a number of novel applications, ranging from index modification writing, high precision welding to high speed cutting of transparent materials. 

A number of efforts are still required to understand the physical processes generating the cavities. This is particularly relevant for silicon. The propagation-invariant properties of Bessel beams are fundamentally at the origin of the possibility to homogeneously deposit energy inside transparent materials with high aspect ratio. We expect that other beam shapes, which are also propagation invariant in the nonlinear regime, will be very attractive in the future to process materials with other geometries and develop novel applications of high-intensity ultrashort pulses.

\bibliographystyle{apalike} 
\bibliography{BookChapterBiblio_current}


\end{document}